\documentclass[twocolumn,floatfix,amsmath,showpacs,amssymb,superscriptaddress,prl]{revtex4}

\usepackage{graphicx}
\usepackage{color}
\usepackage{bm}

\newcommand{\beq}{\begin{equation}}
\newcommand{\eeq}{\end{equation}}
\newcommand{\beqa}{\begin{eqnarray}}
\newcommand{\eeqa}{\end{eqnarray}}

\begin{document}

\title{Correlation-Dependent Coherent to Incoherent Transitions in Resonant Energy Transfer Dynamics}
\author{Ahsan Nazir} \email{ahsan.nazir@ucl.ac.uk}
\affiliation{Department of Physics and Astronomy, University College London, Gower Street, London WC1E 6BT, United Kingdom}
\affiliation{Centre for Quantum Dynamics, Griffith University, Brisbane, Queensland 4111,
Australia}

\date{\today}

\begin{abstract}
I investigate energy transfer in a donor-acceptor pair beyond 
weak system-bath coupling.
I identify a transition from coherent to incoherent  
dynamics with increasing temperature, due to multi-phonon effects not captured by a standard weak-coupling treatment. The  
crossover temperature has a marked dependence on the degree of spatial correlation between fluctuations experienced at the two system sites. For strong correlations, this leads to the possibility of coherence  
surviving into a high temperature regime.
\end{abstract}


\maketitle

Excitation energy transfer is a fundamental process common to a wide variety of multi-site (donor-acceptor) systems, ranging from those in the solid-state, such as crystal impurities~\cite{Foerster59, Soules71,Rackovsky73} and quantum dots (QDs)~\cite{Crooker02,Gerardot05,Kim08}, to conjugated polymers~\cite{Collini09} and photosynthetic complexes~\cite{Renger01,vanGrondelle06,Lee07,Engel07,Cheng09}. In its simplest F\"orster-Dexter (FD) form energy transfer is considered to be incoherent, resulting from weak donor-acceptor transition-dipole interactions~\cite{Foerster59}. However, recent experimental progress in demonstrating {\it quantum coherent} energy transfer in a number of systems~\cite{Lee07,Engel07,Collini09} has highlighted the importance of describing transfer dynamics beyond the incoherent  
regime~\cite{Cheng09}. Furthermore, such systems are still embedded in a larger host matrix, and therefore remain susceptible to couplings to their environment~\cite{Breuer02}. The resulting interplay between coherent and incoherent processes can fundamentally alter the nature of the transfer dynamics, destroying quantum coherent effects and modifying the transfer rate. 

To develop a full 
understanding of any donor-acceptor system it is thus crucial to establish the coherent or incoherent nature of the transfer process~\cite{Rackovsky73,Gilmore06,Leegwater96}, and to explore how this changes with variations in donor-acceptor separation, system-bath coupling strengths, or temperature.
For example, the recent demonstration of coherent transfer at room temperature in conjugated polymers~\cite{Collini09} points to the potentially pivotal role played by correlated dephasing fluctuations in protecting coherence in these systems~\cite{Lee07,Collini09,Yu08}. 
Furthermore, determining the respective roles of coherent and incoherent processes in optimising energy transfer efficiency in donor-acceptor networks is currently subject to considerable interest~\cite{Cheng06,OlayaCastro08,Mohseni08}.

A number of methods have been put forward to deal with the dynamics of coherent energy transfer under the influence of an external environment. A popular assumption is that the system-bath coupling is weak 
\cite{Mohseni08,Rozbicki08}, which 
leads to Redfield-type dynamics involving only single-phonon processes~\cite{Yang02}. A modified Redfield treatment, with a broader range of validity, has also been suggested~\cite{Zhang98,Yang02}. 
For strong system-bath coupling, 
FD theory has been extended to account for exciton delocalisation over donor and acceptor sites~\cite{Sumi99}, while the polaron transformation provides a useful tool to investigate both the weak and strong 
coupling regimes~\cite{Rackovsky73,Wurger98,Jang08}. The importance of 
non-Markovian effects has also been studied~\cite{Kenkre74,Renger01}.

To explore the criteria for coherent energy transfer in a donor-acceptor pair in detail, I present here an analytical theory of the transfer dynamics capable of interpolating between the 
weak (single-phonon) and strong (multi-phonon) system-bath coupling regimes, and correlated to independent fluctuations, while still capturing the coherent dynamics due to the donor-acceptor electronic coupling. As a main result, I identify a crossover from coherent to incoherent transfer for resonant donor-acceptor pairs with increasing temperature, as multi-phonon effects become dominant. Such behaviour cannot be derived from a weak-coupling treatment. 
I show that the critical temperature  at which the crossover occurs has a pronounced dependence on the degree of correlation between fluctuations at each site, leading to the possibility of 
coherent transfer surviving at high temperatures in strongly correlated environments, where multi-phonon processes are suppressed.

Consider a pair of two-level systems ($j=1,2$) separated by a distance ${\bf d}$, with energy 
transfer interaction $V_F$, coupled linearly 
to a harmonic environment 
($\hbar=1$): 
\begin{eqnarray}\label{eqn:sysbathham}
H&{}={}&\sum_{j=1,2}\epsilon_j|X\rangle_j\langle X|+V_F(|GX\rangle\langle XG|+|XG\rangle\langle GX|)\nonumber\\
&&\:{+}\sum_{\bf k}\omega_{\bf k}b_{\bf k}^{\dagger}b_{\bf k}+\sum_{j=1,2}|X\rangle_j\langle X|\sum_{\bf k}(g^{j}_{\bf k}b_{\bf k}^{\dagger}+{g^{j*}_{\bf k}}b_{\bf k}).\nonumber
\end{eqnarray}
Here, each system has ground (excited) state $|G\rangle_j$ ($|X\rangle_j$) and energy $\epsilon_j$, the system-bath couplings are given by $g_{\bf k}^j$, and the bath comprises a collection of oscillators of frequencies $\omega_{\bf k}$ and creation (annihilation) operators $b_{\bf k}^{\dagger}$ ($b_{\bf k}$). Such a model has previously been employed in a range of physical settings, see e.g. Refs.~\cite{Soules71,Rackovsky73,Renger01,Cheng09,Gilmore06,Rozbicki08}, and could also represent the basic unit of a spin chain~\cite{Sinaysky08}. 
We shall consider system-bath couplings of the form $g_{k}^1=|g_{\bf k}|e^{i{\bf k}\cdot\bf{d}/2}$ and $g_{k}^2=|g_{\bf k}|e^{-i{\bf k}\cdot\bf{d}/2}$, where position dependent phases give rise to correlations between the bath-induced fluctuations experienced at each system site~\cite{Soules71,Rackovsky73,Rozbicki08}. Following Refs.~\cite{Rozbicki08,Govorov05}, I parameterise the transfer interaction as $V_F=[V_0/(d/d_0)^3]f(d/d_0)$, where $d=|{\bf d}|$, $f(x)=3\sqrt{\pi/2}\;{\rm erf}(x/\sqrt{2})-3xe^{-x^2/2}$ accounts for small $d$ corrections to the dipole approximation, and $d_0$ determines when the dipole limit is reached. 
 
The full Hamiltonian may be decomposed into three decoupled subspaces [$|GG\rangle,\{|XG\rangle,|GX\rangle\},|XX\rangle$]. We are interested in the energy transfer dynamics occurring between the single-excitation states, described by a Hamiltonian $H_{\rm sub}$, and we set $|XG\rangle\equiv|0\rangle$, $|GX\rangle\equiv|1\rangle$ to identitfy an effective two-state system spanning the subspace~\cite{Gilmore06}. To move into an appropriate basis for the subsequent perturbation theory, we apply the unitary transformation $H'=e^{S}H_{\rm sub}e^{-S}$, where $S=|0\rangle\langle 0|\sum_{\bf k}(\alpha_{\bf k}b_{\bf k}^{\dagger}-\alpha_{\bf k}^*b_{\bf k})+|1\rangle\langle 1|\sum_{\bf k}(\beta_{\bf k}b_{\bf k}^{\dagger}-\beta_{\bf k}^*b_{\bf k})$, with $\alpha_{\bf k}=g^1_{\bf k}/\omega_{\bf k}$ and $\beta_{\bf k}=g^2_{\bf k}/\omega_{\bf k}$. As a result, we map our system to the polaron-transformed, spin-boson model
$H'=\frac{\epsilon}{2}\sigma_z+V_R\sigma_x+\sum_{\bf k}\omega_{\bf k}b_{\bf k}^{\dagger}b_{\bf k}+V_F\left(\sigma_xB_x+\sigma_yB_y\right)$, here describing the energy transfer dynamics of our donor-acceptor pair in the single-excitation subspace, with bath-renormalised coupling $V_R=BV_F$~\cite{Rackovsky73,Wurger98,Jang08}. The Pauli matrices, $\sigma_l$ (for $l=x,y,z$), are defined in the basis $\{|0\rangle,|1\rangle\}$, while $\epsilon=\epsilon_1-\epsilon_2$. Bath-induced fluctuations are described by 
$B_{x}=(1/2)(B_++B_--2B)$ and $B_{y}=(-i/2)(B_--B_+)$, where $B_{\pm}=\Pi_{\bf k}D(\pm(\alpha_{\bf k}-\beta_{\bf k}))$ are products of displacement operators $D(\pm\chi_{\bf k})=e^{\pm(\chi_{\bf k}b_{\bf k}^{\dagger}-\chi_{\bf k}^*b_{\bf k})}$~\cite{Wurger98}. Assuming the bath to be in thermal equilibrium at temperature $T$, the correlation-dependent renormalisation of the coupling strength is determined by $B=\langle B_{\pm}\rangle=e^{-\int_0^{\infty}d\omega\frac{J(\omega)}{\omega^2}(1-F(\omega,d))\coth{(\beta\omega/2)}}$, where $\beta=1/k_BT$, with Boltzmann constant $k_B$. Here, we define  a single-site spectral density as $J(\omega)=\sum_{\bf k}|g_{\bf k}|^2\delta(\omega-\omega_{\bf k})$, while the factor  $\left(1-F(\omega,d)\right)$ accounts for the degree of spatial correlation in the fluctuations at each site. We find $F(\omega,d)={\rm sinc}{(\omega d/c)}$ in $3$-dimensions, assuming $k=\omega/c$, and that $J(\omega)$ is isotropic. 
 
We now write $H'=H'_{0}+H'_{I}$, where $H'_{I}=V_F\left(\sigma_xB_x+\sigma_yB_y\right)$ is treated as a perturbation. Provided $V_R$ is non-zero, as assumed throughout, this procedure is suitable for exploring both single- and multi-phonon bath-induced effects on the system dynamics. In cases where $V_R=0$, we can instead apply a related variational approach~\cite{Silbey84}. Following the standard procedure~\cite{epaps}, we derive a Markovian master equation describing the reduced system dynamics in the polaron-transformed Schr\"odinger picture (${\rm h.c.}$ denotes the Hermitian conjugate)~\cite{Breuer02}:
\begin{equation}\label{eqn:schpictmastereigen}
\dot{\rho'}=-\frac{i\eta}{2}[\sigma_z,\rho']-V_F^2\sum_{l,\omega,\omega'}(\Lambda_l(\omega')[P_l(\omega),P_l(\omega')\rho']+{\rm h.c.}),
\end{equation}
where $\omega,\omega'\in\{0,\pm\eta\}$, $\eta=\sqrt{\epsilon^2+4V_R^2}$, $\Lambda_l(\omega)=\gamma_l(\omega)/2+iS_l(\omega)$, and we have decomposed the  
system operators as $\hat{\sigma}_l(t)=\sum_{\omega}P_l(\omega)e^{-i\omega t}$~\cite{epaps}. 
The rates
\begin{equation}\label{rates}
\gamma_l(\omega)=e^{\beta\omega/2}\int_{-\infty}^{\infty}d\tau e^{i\omega\tau}\langle B_l^{\dagger}(\tau-i\beta/2)B_l(0)\rangle,
\end{equation}
are Fourier transforms of the bath correlation functions
\begin{eqnarray} 
\langle B_x^{\dagger}(\tau-i\beta/2)B_x(0)\rangle&=&(B^2/2)(e^{\bar{\varphi}(\tau)}+e^{-\bar{\varphi}(\tau)}-2),\;\;\;\label{bxcorr}\\
\langle B_y^{\dagger}(\tau-i\beta/2)B_y(0)\rangle&=&(B^2/2)(e^{\bar{\varphi}(\tau)}-e^{-\bar{\varphi}(\tau)}),\label{bycorr}
\end{eqnarray} 
defined in terms of the phonon propagator~\cite{Wurger98}
\begin{eqnarray}\label{eqn:phononpropspecdens}
\bar{\varphi}(\tau)=&{}2{}&\int_0^{\infty}d\omega\frac{J(\omega)}{\omega^2}(1-F(\omega,d))\frac{\cos{\omega\tau}}{\sinh{(\beta\omega/2)}},
\end{eqnarray}
while $S_l(\omega)={\rm Im}\int_0^{\infty}d\tau e^{i\omega\tau}\langle B_l^{\dagger}(\tau)B_l(0)\rangle$. The most interesting dynamics of the model can now be explored by considering two limiting cases: that of resonant donor and acceptor, in which the interplay between coherent and incoherent processes is most evident, and that of large energy mismatch, often encountered in practise. 

{\it Resonant} -  
The resonant case is of particular importance as it demonstrates most clearly how bath-induced fluctuations can fundamentally alter the nature of the energy transfer process. As we shall see, in the high-temperature regime, multi-phonon dephasing effects can become dominant, giving rise to a crossover from low-temperature coherent dynamics to a high-temperature incoherent process. Setting $\epsilon=0$, we derive from Eq.~(\ref{eqn:schpictmastereigen}) a set of Bloch equations governing the time evolution of the system state. Taking an initial state $\rho'(0)=|XG\rangle\langle XG|$, a single excitation in the donor, and transforming out of the polaron frame, we solve for the subsequent donor-acceptor population dynamics, $\langle\sigma_z\rangle_t={\rm Tr}_S(\sigma_z\rho(t))$, to find 
\begin{equation}
\langle\sigma_z\rangle_t=e^{-(\Gamma_1+\Gamma_2)t/2}\left(\cos{\frac{\xi t}{2}}+\frac{(\Gamma_2-\Gamma_1)}{\xi}\sin{\frac{\xi t}{2}}\right),\label{szresonant}
\end{equation}
where $\xi=\sqrt{8V_R(2V_R+\lambda)-(\Gamma_1-\Gamma_2)^2}$. Here, 
\begin{eqnarray}
\Gamma_1&{}={}&V_F^2\left[2\gamma_x(0)+\gamma_y(2V_R)\frac{(1+2N(2V_R))}{(1+N(2V_R))}\right],\\
\Gamma_2&{}={}&2V_F^2\gamma_x(0),
\end{eqnarray} 
$\lambda=2V_F^2\left(S_y(2V_R)-S_y(-2V_R)\right)$, and $N(\omega)=(e^{\beta\omega}-1)^{-1}$. The coherent-incoherent transition thus occurs at 
\begin{equation}\label{crossovercondition}
8V_R(2V_R+\lambda)=(\Gamma_1-\Gamma_2)^2.
\end{equation}

We shall return to the crossover shortly. First, let's consider the dynamics in the weak system-bath coupling limit. In this case, we expand Eqs.~(\ref{bxcorr}) and~(\ref{bycorr}) to first order in $\bar{\varphi}(\tau)$,  
hence keeping only single-phonon contributions. We then find $\Gamma_2\approx0$, and thus a damping rate $\tilde{\Gamma}_1=\pi J(2\tilde{V}_R)(1-F(2\tilde{V}_R,d))\coth{\beta\tilde{V}_R}$. Here, $\tilde{V}_R=\tilde{B}V_F$, where we expand $\tilde{B}\approx B_0[1-\int_0^{\infty}d\omega\frac{J(\omega)}{\omega^2}(1-F(\omega,d))(\coth{\beta\omega/2}-1)]$, with vacuum term $B_0=e^{-\int_0^{\infty}d\omega\frac{J(\omega)}{\omega^2}(1-F(\omega,d))}$. From Eq.~(\ref{szresonant}), we find that the system performs damped coherent oscillations: $\langle\sigma_z\rangle_t=e^{-\tilde{\Gamma}_1t/2}[\cos{(\tilde{\xi} t/2)}-(\tilde{\Gamma}_1/\tilde{\xi})\sin{(\tilde{\xi} t/2)}]$, with frequency $\tilde{\xi}\approx\sqrt{16\tilde{V}_R^2-\tilde{\Gamma}_1^2}$, where $4\tilde{V}_R>\tilde{\Gamma}_1$ to be consistent with the original expansion. 

Under what circumstances is a weak-coupling approximation appropriate? To address this question it is useful to consider an explicit form for the spectral density. As an illustration, we choose $J(\omega)=A\omega^3$, describing, for example, acoustic phonon induced dephasing with a coupling strength $A$~\cite{Wurger98,Rozbicki08,Nazir08}. Here, we keep a cutoff frequency $\omega_c$ only in the vacuum terms. From Eq.~(\ref{eqn:phononpropspecdens}) we obtain
\begin{equation}\label{eqn:phibarsuperohmic3D}
\bar{\varphi}(\tau')=\varphi_0\left({\rm sech}^2\tau'-\frac{\tanh{(x-\tau')}+\tanh{(x+\tau')}}{2x}\right),
\end{equation}
where we scale the time as $\tau'=\pi\tau/\beta$, and define the dimensionless parameters $\varphi_0=2\pi^2A/\beta^2=T^2/T_0^2$ and $x=\pi d/c\beta=T/T_d$. Importantly, we can now identify two distinct temperature scales that determine whether single-phonon or multi-phonon processes are relevant: $T_0$, set by $A$~\cite{Wurger98}; and $T_d$, which is inversely proportional to the separation, 
and is therefore {\it correlation-dependent}. 

\begin{figure}[!t]
\centering
\includegraphics[width=0.37\textwidth]{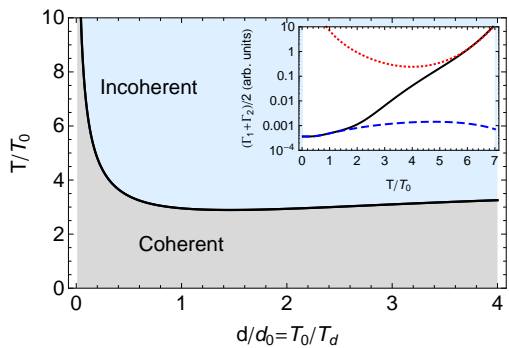}
\caption{(Color online) Main: Regimes of resonant energy transfer for varying temperature ($T/T_0$) and scaled donor-acceptor separation. The line $T=T_c$, given by Eq.~(\ref{crossovertemp}), divides the coherent (lower) and incoherent (upper) cases. Inset: Resonant damping rate versus $T/T_0$ evaluated numerically (black, solid line), and by single-phonon (blue, dashed line) and high-temperature (red, dotted line) analytical approximations. Here, $T_d/T_0=10$. Parameters: $\omega_c/k_BT_0=5$ and $V_0/k_BT_0=1$.} 
\label{tstar}
\end{figure}

Let's consider two cases: (i) when $x\gg1$ ($T\gg T_d$, weak fluctuation correlations), it can be shown from Eq.~(\ref{eqn:phibarsuperohmic3D}) that $\varphi_0$ alone is suitable as an expansion parameter in the bath correlation functions. Hence, $\varphi_0\ll 1$ defines the single-phonon regime in this case, most easily satisfied for large separation $d$, small $A$, and low $T$; (ii) when $x\ll1$, the strongly-correlated case most easily satisfied for small $d$, we expand Eq.~(\ref{eqn:phibarsuperohmic3D}) to second-order in $x$ to give $\bar{\varphi}(\tau)\approx\varphi_0x^2((1-4\tanh^2{\tau'})/3+\tanh^4{\tau'})$. Now, $\varphi_0x^2$ plays the role of an expansion parameter in the correlation functions, with the single-phonon rate valid for $\varphi_0x^2\ll1$. However, since $x$ is already assumed small in this case, it is clear that the single-phonon rate can be used at least up to $\varphi_0\sim1$, and is therefore valid over a much larger range of temperatures and/or coupling strengths than in case (i). The system is thus far better protected from the adverse effects of the environment when the fluctuations are highly correlated, and hence multi-phonon processes can be suppressed up to much higher temperatures. This 
is shown in the inset of Fig.~\ref{tstar}, where the damping rate in Eq.~(\ref{szresonant}) is plotted against temperature for strong correlations, leading to a single-phonon rate valid beyond $T/T_0=1$.

Turning now to the high-temperature 
regime, the rates are estimated by expanding $\bar{\varphi}(\tau)$ about $\tau=0$, where it is strongly peaked. Keeping terms up to $\tau^2$ order 
we find 
\begin{equation}\label{highTratesresonance}
\Gamma_1\approx2\Gamma_2\approx2\beta\frac{V_F^2B_0^2e^{2\varphi_0/3}e^{\varphi_0(2x{\rm csch}{2x}-1)/x^2}}{\sqrt{\pi\varphi_0(x-{\rm sech}^2x\tanh{x})/x}},
\end{equation}
valid for 
$2\beta V_R<1$, with $B_0^2=e^{-\Omega^4x^2/(\varphi_0+\Omega^2x^2)}$, where $\Omega=\omega_c/\pi k_BT_0$. Further, $\lambda\approx0$ in this limit, hence $\xi\approx\sqrt{16V_R^2-\Gamma_1^2/4}\approx i\Gamma_1/2$ in Eq.~(\ref{szresonant}), giving $\langle\sigma_z\rangle_t\approx e^{-\Gamma_1t}$. Thus, in the high-temperature resonant case, the transfer is incoherent, 
at a rate $\Gamma_1$ given in Eq.~(\ref{highTratesresonance}).

The transition between these two regimes, from coherent to incoherent dynamics, is particularly important as it allows us to assess up to what critical temperature quantum coherent effects might be observed. As we have seen, the weak-coupling dynamics is expected to be coherent, hence the crossover generally occurs in the high-temperature regime, where Eq.~(\ref{highTratesresonance}) is valid. Then, Eq.~(\ref{crossovercondition}) simplifies to $8V_R=\Gamma_1$, with the  
transfer 
being coherent for $8V_R>\Gamma_1$. We use this condition to define a critical temperature, $T_c$, above which the dynamics becomes incoherent. From Eq.~(\ref{highTratesresonance}) we find the implicit equation
\begin{equation}\label{crossovertemp}
T_c^2=T_0\frac{V_FB_0e^{5\varphi_c/6}e^{\varphi_c(\coth{x_c}-2\tanh{x_c}-1/x_c)/2x_c}}{4k_B\sqrt{\pi(x_c-{\rm sech}^2x_c\tanh{x_c})/x_c}},
\end{equation}
where $\varphi_c=T_c^2/T_0^2$ and $x_c=T_c/T_d$. It is clear that $T_c$ will vary in a nontrivial way as a function of donor-acceptor separation, through the dependence of Eq.~(\ref{crossovertemp}) on $x_c$, $B_0$, and $V_F$. Again, we consider two limits: (i) as the separation becomes large, the ``correlation" temperature becomes unimportant ($T_d\rightarrow0$) 
and $T_c$ varies only weakly with separation through $V_F$; 
(ii) at very small separations the rates $\Gamma_1$ and $\Gamma_2$ tend to zero, while $V_R\rightarrow V_F$. Hence, in this limit, $T_c$ diverges, as we expect; for 
complete fluctuation correlation the system behaviour is always coherent, with no crossover to incoherent dynamics regardless of the temperature. 

To illustrate this behaviour, in the main part of Fig.~\ref{tstar} we plot the crossover temperature, shown separating the coherent and incoherent regimes, as a function of donor-acceptor separation. The divergence of $T_c$ at small $d$ implies that coherent dynamics can survive at elevated temperatures when strong fluctuation correlations suppress multi-phonon effects, consistent with recent experimental observations~\cite{Collini09}. Further, the change in $T_c$ behaviour from small to large separations can provide information on the correlation length of the bath.  
Specifically, once the distance dependence of $T_c$ becomes weak  
there is no longer significant correlation between  
fluctuations  
at each site. 

To give Fig.~\ref{tstar} a relevant experimental context, we now estimate $T_0$ and $T_d$ for two closely-spaced semiconductor QDs, as realized experimentally in Ref.~\cite{Gerardot05}, which could be brought into resonance by applying an external electric field. Typically, deformation potential coupling to acoustic phonons dominates exciton dephasing in such samples~\cite{Rozbicki08}. A simple model~\cite{Nazir08} allows an estimate of $A=0.032$~ps$^2$ in this case, implying $T_0\approx9.6$~K. Taking $c=5110$~ms$^{-1}$~\cite{Rozbicki08} and a dot separation $d=4.5$~nm~\cite{Gerardot05}, we find $T_d\approx2.8$~K. Setting $V_0/k_BT_0=1$ and $d/d_0=T_0/T_d$ then implies the reasonable values $V_0\approx0.8$~meV and $d_0\approx1.3$~nm~\cite{Rozbicki08}, respectively. From Fig.~\ref{tstar} we then obtain a crossover temperature of $T_c\approx30$~K, below which we expect the energy transfer dynamics to display signatures of coherence. In fact, in Ref.~\cite{Gerardot05} temperatures of around $4-40$~K were explored, which should therefore be a promising range over which to observe both coherent and incoherent transfer dynamics in QD samples.

{\it{Off-resonant}} - 
It is also important to examine the dynamics when the donor and acceptor are far off-resonant with each other, such that $V_F/\epsilon\ll1$. This can occur quite naturally, for example in QD samples due to the nature of their growth. Furthermore, the recent weak-coupling 
theory of Ref.~\cite{Rozbicki08} predicts a single transfer rate in the off-resonant regime, and thus provides a means to assess the validity of our theory in this limit.  As in the resonant case, we derive a set of Bloch equations from Eq.~(\ref{eqn:schpictmastereigen}), this time expanding the resulting expressions to second-order in $V_F/\epsilon$. We find system dynamics well approximated by $\langle\sigma_z\rangle_t\approx e^{-\Gamma t}-(1-e^{-\Gamma t})\tanh{(\beta\epsilon/2)}$, describing incoherent energy transfer from the initially excited donor to the acceptor at a rate $\Gamma= V_F^2\frac{(1+2N(\epsilon))}{(1+N(\epsilon))}(\gamma_x(\epsilon)+\gamma_y(\epsilon))$. Taking the weak coupling limit of $\Gamma$ by retaining only single-phonon terms, we find $\tilde{\Gamma}\approx (4\pi\tilde{V}_R^2/\epsilon^2)J(\epsilon)(1-F(\epsilon,d))\coth{(\beta\epsilon/2)}$, consistent with Ref.~\cite{Rozbicki08} once renormalisation of $V_R$ has been included there. In the opposite, high temperature limit ($k_BT\gg\epsilon$), we again find $\langle\sigma_z\rangle_t\approx e^{-\Gamma t}$, with $\Gamma=\Gamma_1$ of Eq.~(\ref{highTratesresonance}).

{\it{Summary}} - 
I have presented an analytical theory of excitation transfer in a correlated environment, showing that for resonant donor and acceptor, a crossover from coherent to incoherent transfer is expected as multi-phonon effects begin to dominate. 
The theory outlined here opens up intriguing possibilities for further study of the role of coherence in the transfer dynamics of larger arrays, such as photosynthetic complexes~\cite{Cheng06,OlayaCastro08,Mohseni08}. For example, it enables one to address the important question of how the transfer efficiency changes in such systems when crossing from the coherent to incoherent regime.

I am very grateful to A. Olaya-Castro, S. Bose, and A. M. Stoneham for useful discussions. I am supported by the EPSRC, Griffith University, and the Australian Research Council Centre for Quantum Computer Technology.

\widetext

\newpage

\section{Supplement: Derivation of the master equation}

{\small
In this supplement, I outline the derivation of the master equation (Eq. (1) of the paper) used to describe the donor-acceptor energy transfer dynamics in the single-excitation subspace.}\\

As in the paper, we write the polaron-transformed Hamiltonian within the single excitation subspace $\{|XG\rangle,|GX\rangle \}$ as
\begin{equation}
\label{Hsubpol}
H'=\frac{\epsilon}{2}\sigma_z+V_R\sigma_x+\sum_{\bf k}\omega_{\bf k}b_{\bf k}^{\dagger}b_{\bf k}+V_F\left(\sigma_xB_x+\sigma_yB_y\right),
\end{equation}
where we set $|XG\rangle\equiv|0\rangle$, $|GX\rangle\equiv|1\rangle$, and define $\sigma_z=|0\rangle\langle0|-|1\rangle\langle1|$, $\sigma_x=\sigma_++\sigma_-=|0\rangle\langle1|+|1\rangle\langle0|$, and $\sigma_y=i(\sigma_--\sigma_+)=i(|1\rangle\langle0|-|0\rangle\langle1|)$.
We now separate the Hamiltonian as $H'=H'_{0}+H'_{I}$, where $H'_0=H_S'+H_B'$, with
\begin{eqnarray}\label{H0pol}
H_S'&{}={}&\frac{\epsilon}{2}\sigma_z+V_R\sigma_x,\\
H_B'&{}={}&\sum_{\bf k}\omega_{\bf k}b_{\bf k}^{\dagger}b_{\bf k},
\end{eqnarray}
while
\begin{equation}\label{HIpol}
H'_{I}=V_F\left(\sigma_xB_x+\sigma_yB_y\right),
\end{equation}
will be treated as a perturbation. 

To proceed with deriving the master equation, we first diagonalise the system part of the polaron-transformed Hamiltonian, $H_{S}'$, by applying the rotation $U^{\dagger}H'U$, where $U=e^{-i\vartheta\sigma_y/2}$ and $\vartheta=\arctan{(2V_R/\epsilon)}$. This gives
\begin{equation}\label{eqn:HPoldiag}
U^{\dagger}H'U=\frac{\eta}{2}\sigma_z+\sum_{\bf k}\omega_{\bf k}b^{\dagger}_{\bf k}b_{\bf k}+V_F\left[(\sin{\vartheta}\sigma_z+\cos{\vartheta}\sigma_x)B_x+\sigma_yB_y\right],
\end{equation} 
where $\eta=\sqrt{\epsilon^2+4V_R^2}$, $\cos{\vartheta}=\epsilon/\eta$, and $\sin{\vartheta}=2V_R/\eta$. Now, moving into the interaction picture with respect to $(\eta/2)\sigma_z+\sum_{\bf k}\omega_{\bf k}b^{\dagger}_{\bf k}b_{\bf k}$, we obtain an interaction Hamiltonian of the form 
\begin{equation}\label{eqn:Hpolint}
H_{I}'(t)=V_F(\hat{\sigma}_x(t)B_x(t)+\hat{\sigma}_y(t)B_y(t)),
\end{equation}
where we decompose the system operators  
as~\cite{Breuersupp02}
\begin{equation}\label{sigmadecomp}
\hat{\sigma}_l(t)=\sum_{\omega\in\{0,\pm\eta\}}P_l(\omega)e^{-i\omega t},
\end{equation} 
where $l=x,y$ and
\begin{eqnarray}\label{eigenops}
P_x(0)&{}={}&\sin{\vartheta}\sigma_z,\nonumber\\
P_x(\pm\eta)&{}={}&\cos{\vartheta}\sigma_{\mp},\nonumber\\
P_y(0)&{}={}&0,\nonumber\\
P_y(\pm\eta)&{}={}&\pm i\sigma_{\mp}.
\end{eqnarray}
Furthermore, the bath operators transform as 
\begin{eqnarray}\label{Bt}
B_x(t)&{}={}&\frac{1}{2}(B_+(t)+B_-(t)-2B),\label{Btx}\\
B_y(t)&{}={}&\frac{1}{2i}(B_-(t)-B_+(t)),\label{Bty}
\end{eqnarray} 
where 
\begin{equation}\label{binteraction}
B_{\pm}(t)=\Pi_{\bf k}D(\pm(\alpha_{\bf k}-\beta_{\bf k})e^{i\omega_{\bf k}t}),
\end{equation} 
are written in terms of the (now time-dependent) bath displacement operators $D(\pm\chi_{\bf k})=e^{\pm(\chi_{\bf k}b_{\bf k}^{\dagger}-\chi_{\bf k}^*b_{\bf k})}$.

We now follow the standard procedure to derive a Markovian master equation~\cite{Breuersupp02}, here governing the dynamics of the reduced system density operator $\rho'(t)$ in the polaron frame. We integrate the von Neumann equation for the joint system-bath density operator in the polaron frame interaction picture, $\chi_I'(t)$, then trace over the bath modes. This results in an integro-differential equation for the reduced density operator within the interation picture of the form
\begin{equation}\label{integrodiff}
\frac{d\rho'_I(t)}{dt}=-\int_0^tds{\rm Tr}_B\left(\big[H_I'(t),\big[H_I'(s),\chi_I'(s)\big]\big]\right),
\end{equation}
where we assume factorising initial conditions, $\chi'(0)=\rho'(0)\rho_B$, with $\rho_B=e^{-H_B'/k_BT}/{\rm Tr}_B(e^{-H_B'/k_BT})$ being the thermal equilibrium state of the bath, and use ${\rm Tr}_B(H_I'(t)\rho_B)=0$. To perform the Born-Markov approximation, we now make two assumptions. First, that the perturbation of the bath state is weak during the combined system-bath evolution, so that we may factorize the joint density operator as $\chi'_I(t)=\rho'_I(t)\rho_B$ at all times. Second, that the timescale on which the system evolves appreciably is large compared to the bath memory time $\tau_B$, allowing us to replace $\rho_I'(s)$ by $\rho_I'(t)$ in Eq.~(\ref{integrodiff}), giving
\begin{equation}\label{1stMarkov}
\frac{d\rho'_I(t)}{dt}=-\int_0^tds{\rm Tr}_B\left(\big[H_I'(t),\big[H_I'(s),\rho_I'(t)\rho_B\big]\big]\right).
\end{equation}
To complete the Markov approximation, we make a change of variable $t-s\rightarrow\tau$, and take the upper limit of integration to infinity, to give
\begin{equation}\label{Markovschrodinger}
\frac{d\rho_I'(t)}{dt}=-\int_0^{\infty}d\tau{\rm Tr}_B\left(\big[H_I'(t),\big[H_I'(t-\tau),\rho_I'(t)\rho_B\big]\big]\right).
\end{equation}
Substituting Eq.~(\ref{eqn:Hpolint}) into Eq.~(\ref{Markovschrodinger}), transforming out of the interaction picture, and using Eqs.~(\ref{sigmadecomp})~-~(\ref{binteraction}), leads directly to the Markovian master equation given in Eq.~(1) of the paper,
\begin{equation}\label{eqn:schpictmastereigen}
\frac{d\rho'(t)}{dt}=-\frac{i\eta}{2}[\sigma_z,\rho'(t)]-V_F^2\sum_{l,\omega,\omega'}\left(\Lambda_l(\omega')[P_l(\omega),P_l(\omega')\rho'(t)]+{\rm H.c.}\right),
\end{equation}
describing the polaron-transformed Scr\"odinger picture dynamics on timescales $\tau_S>\tau_B$, where $\tau_B\sim1/\omega_c$ at $T=0$, with $\omega_c$ being a high-frequency cutoff in the bath spectral density. Here, $\omega,\omega'\in\{0,\pm\eta\}$, H.c. denotes the Hermitian conjugate, while $\Lambda_l(\omega)$ are one-sided Fourier transforms of the bath correlation functions:
\begin{equation}\label{eqn:ftbath}
\Lambda_l(\omega)=\int_0^{\infty}d\tau e^{i\omega\tau}\langle B_l^{\dagger}(\tau)B_l(0)\rangle,
\end{equation}
where $\langle B_l^{\dagger}(\tau)B_l(0)\rangle\equiv{\rm Tr}_B(B_l^{\dagger}(\tau)B_l(0)\rho_B)$ are evaluated using Eqs.~(\ref{Btx}),~(\ref{Bty}), and~(\ref{binteraction}). Setting $\Lambda_l(\omega)=\gamma_l(\omega)/2+iS_l(\omega)$ to separate real and imaginary parts, we see that the rates $\gamma_l(\omega)$ may be written precisely as in Eq.~(2) of the paper,
\begin{eqnarray}\label{eqn:gammal}
\gamma_l(\omega)=\Lambda_l(\omega)+\Lambda_l(\omega)^*&{}={}&\int_{-\infty}^{\infty}d\tau e^{i\omega\tau}\langle B_l^{\dagger}(\tau)B_l(0)\rangle,\nonumber\\
&{}={}&e^{\beta\omega/2}\int_{-\infty}^{\infty}d\tau e^{i\omega\tau}\langle B_l^{\dagger}(\tau-i\beta/2)B_l(0)\rangle,
\end{eqnarray}
where a change of integration variable $\tau\rightarrow\tau-i\beta/2$ has been made in the second line~\cite{Wurgersupp98}.

\bibliographystyle{apsrev}

\end{document}